\documentclass[twocolumn,pra,floatfix,longbibliography,superscriptaddress,showpacs]{revtex4-1}

\usepackage{amsmath}
\usepackage{amsfonts}
\usepackage{amsbsy}
\usepackage{xcolor}
\usepackage{soul}
\usepackage{graphicx}
\usepackage{epsfig}
\usepackage{hyperref}

\newcommand{\rv}{{\bf r}}
\newcommand{\Ev}{{\bf E}}
\newcommand{\Dv}{{\bf D}}
\newcommand{\Pv}{{\bf P}}

\newcommand{\eo}{\epsilon_0}
\newcommand{\beq}{\begin{equation}}
\newcommand{\eeq}{\end{equation}}
\newcommand{\bea}{\begin{eqnarray}}
\newcommand{\eea}{\end{eqnarray}}
\newcommand{\BEQAL}{\begin{align}}
\newcommand{\EEQAL}{\end{align}}
\newcommand{\EQREF}[1]{Eq.~(\ref{#1})}
\newcommand{\eq}[1]{(\ref{#1})}
\newcommand{\comment}[1]{{}}
\newcommand{\half}{\hbox{$1\over2$}}

\renewcommand{\>}{\rangle}

\newcommand{\commentout}[1]{{}}

\begin{document}
\title{Arrays of strongly-coupled  atoms in a one-dimensional waveguide}
\author{Janne Ruostekoski}
\affiliation{Mathematical Sciences, University of Southampton, Southampton SO17 1BJ, United Kingdom}
\author{Juha Javanainen}
\affiliation{Department of Physics, University of Connecticut, Storrs, Connecticut 06269-3046}
\date{\today}
\begin{abstract}
We study the cooperative optical coupling between regularly spaced atoms in a one-dimensional waveguide using decompositions to subradiant and superradiant collective excitation eigenmodes, direct numerical solutions, and analytical transfer-matrix methods. We illustrate how the spectrum of transmitted light through the waveguide including the emergence of narrow Fano resonances can be understood by the resonance features of the eigenmodes.
We describe a method based on superradiant and subradiant modes to engineer the optical response of the waveguide and to store light. 
The stopping of light is obtained by transferring an atomic excitation to a subradiant collective mode with the zero radiative resonance linewidth by controlling the level shift of an atom in the waveguide. 
Moreover, we obtain an exact analytic solution for the transmitted light through the waveguide for the case of a regular lattice of atoms and provide a simple description how
the light transmission may present large resonance shifts when the lattice spacing is close, but not exactly equal, to half of the wavelength of the light.  Experimental imperfections such as fluctuations of the positions of the atoms and loss of light from the waveguide are easily quantified in the numerical simulations, which produce the natural result that the optical response of the atomic array tends toward the response of a gas with random atomic positions.
\end{abstract}
\maketitle

\section{Introduction}

Nanophotonic waveguides confine light and are promising systems to engineer enhanced interactions between light and matter. Hybrid systems, where atoms are integrated with nanophotonic devices, have experienced a rapid development in recent years.  Waveguides~\cite{kimblenaturecom,kimbletrap,kimblesuper} and nanofibers~\cite{Mitsch,Rauschenbeutel,kimblefiber,hakuta1,pfaufiber1,Windpassinger_fiber} closely mimic
one-dimensional light propagation~\cite{Javanainen1999a,Kien09,kien_super,Zubairy,Ruostekoski_waveguide,Manzoni17}
that may have applications in quantum networks, light circuitry, and quantum switches~\cite{Fan1,lukintransistor,Shen_twophoton,Nori,chang_mirror,chang_multiatom,Zhou_atommirrors,Tiecke,Zoubi_Ritsch}, or high-precision spectroscopy~\cite{lambdickespec}. The strong light-mediated interactions between the atoms, and atom-mediated interactions between photons, could be utilized in the preparation of novel quantum many-body systems~\cite{cirackimble,kimblemanybody} for atoms and light.

In our previous study~\cite{Ruostekoski_waveguide} we considered waveguides comprising ensembles of atoms with a constant density. The atoms were either entirely randomly distributed or quantum degenerate. In such systems the transmission of light through the waveguide was shown to display correlation effects even if the atoms are classical, and entirely uncorrelated before the light enters the waveguide. The uniform density distribution mimics a continuous optical medium with a uniform refractive index, yet the light is still able to establish correlations between individual atoms that depend on the atomic positions. Light propagation inside the waveguide then becomes highly nontrivial: light-induced correlations result in the violation of the mean-field-theoretical response~\cite{Javanainen2014a,JavanainenMFT} that is also a basic assumption behind the standard continuous-medium optics~\cite{Jackson,BOR99}.

Here we study a different setting where the atoms are at fixed or weakly fluctuating positions and can also form periodic lattices, using three complementary methods: direct numerical solutions, discussions in terms of the cooperative eigenmodes of the atoms and the electromagnetic field, and analytical transfer-matrix methods. Generally speaking, the optical response dramatically differs from that of a disordered atomic medium. A fixed, regular lattice, for example, displays Bragg resonances that depend on the lattice constant -- a phenomenon that has an analogy in the spectroscopy of crystal lattices.

As in Ref.~\cite{Ruostekoski_waveguide}, we consider strong light-mediated coupling between the atoms, and all the recurrent scattering processes~\cite{Ishimaru1978,Morice1995a,Ruostekoski1997a} -- where light scatters more than once by the same atom -- are fully accounted for. In the limit of low light intensity and for stationary atoms the simulations of the specific 1D electrodynamics are exact and reveal an interplay between superradiant and subradiant eigenmodes of collective atomic excitations. The interference between the broad superradiant and narrow subradiant collective modes can lead to narrow Fano resonances~\cite{Zubairy} in transmission, analogously to the destructive single-atom resonances in the electromagnetically-induced transparency (EIT)~\cite{FleischhauerEtAlRMP2005}. We find that, depending on the interatomic separation and control of the detunings, light can be employed to engineer atomic excitations with nontrivial symmetries. In particular, we show how the collective excitations of the atoms in the waveguide can be used to stop the light in the zero radiative resonance linewidth states. The method is based on the manipulation of the level shift of one atom that drives the excitation to the subradiant mode that otherwise would be entirely decoupled from the incident light.
Furthermore, for a periodic lattice of atoms we provide an exact analytic solution for the transmitted light. This yields the locations of the transmission resonances, and also leads to the prediction that for certain lattice spacing anomalously large shifts of the optical resonance of the waveguide may occur. Finally, we demonstrate how the effects of fluctuations in the atomic positions that result from a weakened confinement of the atoms and the photon losses from the waveguide are easily quantified using numerical simulations.

The interaction of light with many-atom systems in confined geometries and in atom chains has been studied in several related works.
The effects of a photon pulse propagation in a chain of atoms in a waveguide were studied in Ref.~\cite{Zubairy} in the similar linear regime where each atom responds to light as a classical oscillator. Fano-like resonances were identified in a two-atom system and the frequency dependence of the reflectivity was numerically analyzed for the pulse propagation in atom chains. The collective interaction of light with 1D chains of multiple atoms with the effect on resonance linewidths and other collective phenomena has also been studied in free space~\cite{Clemens2003a,Sutherland1D,Bettles1D,Asenjo_prx} and in cavity QED~\cite{Elliott15,Mazzucchi16,Lee17,Genes17}. For incorporating the effects of fluctuations of the atomic positions  we use in this paper the techniques introduced in the simulations of the cooperative response of atoms in optical lattices to the incident light~\cite{Jenkins2012a}.

The paper is organized as follows: In Sec.~\ref{model} we introduce the basic relations for the interactions of light with the many-atom systems, the numerical model, and the description of the collective excitation eigenmodes. The calculations of the transmitted light and the role of the eigenmodes in the spectrum for simple systems is introduced in Sec.~\ref{fewatoms}. The main results for the periodic lattice of atoms are presented in Sec.~\ref{arrays}, including the techniques for stopping light and the analysis of the resonance shifts. The effect of confinement of atoms and the losses of photons from the waveguide are discussed in Sec.~\ref{confinementlosses}. Some concluding remarks are made in Sec.~\ref{conc}. The Appendices provide the description of the transfer matrix methods including the exact analytic solutions for the transmitted light, and further details of the preparation of the subradiant excitations.

\section{Model for 1D electrodynamics}
\label{model}

\subsection{Light and atoms}

We follow our previously developed formalism~\cite{Ruostekoski_waveguide} and describe the propagation of light inside a single-mode waveguide by an effective 1D electrodynamics.
We assume that there is a dominant frequency $\Omega=kc$ of the driving light and, for simplicity of notation, here and in the rest of the paper we have written all operators and classical quantities
in the  ``slowly varying'' picture by explicitly factoring
out the dominant frequency component: $\Dv^+\rightarrow e^{-i\Omega t} \Dv^+$, $ \Pv^+\rightarrow e^{-i\Omega t} \Pv^+$, $ \Ev^+\rightarrow e^{-i\Omega t} \Ev^+$, and so on, where $\Dv^+$, $\Pv^+$, and $\Ev^+$ denote the positive frequency components of the electric displacement, atomic polarization, and electric field, respectively.
For instance, in situations where the atoms are strongly confined close to the center of the waveguide and when an effective two-level system is obtained from the $J=0\rightarrow J'=1$ system, we may obtain 1D scalar electrodynamics for the coupled system of atoms and light by renormalizing the fields, as in $ \pi\xi_\varrho^2 \Ev^+(\rv)\rightarrow \tilde E^+(x)$ and $ \pi\xi_\varrho^2 \Dv^+_F(\rv)\rightarrow \tilde D^+_F(x)$, where $\xi_\varrho$ is the effective radius of the light mode. We also integrate over the radial dependence of the atomic polarization and atom density and, for simplicity of notation, assume that they have the same radial profile.

The total electric field amplitude is the sum of the incident and the scattered fields,
\beq
\eo \tilde E^+(x) = \tilde D^+_F(x) + \sum_l G(x-x_l) \mathfrak{ P}^{(l)}\,,
\label{eq:MonoD1d}
\eeq
where $ D^+_F$ denotes the incident (coherent) field and the scattering from the excitation dipole of the atom $l$, $\mathfrak{ P}^{(l)}$, located at $x_l$,  is described the Green's function of the 1D Helmholtz differential operator,
\beq
G(x)= {ik\over2} e^{ik|x|}\,,
\eeq
with
\begin{equation}
(\nabla^2+k^2)\langle{\tilde D}^+_F\rangle = 0, \quad
(\nabla^2+k^2)\, G(x) = -\delta(x)\,.
\end{equation}
1D Maxwell's wave equation in a polarizable medium~\cite{BOR99} can then be obtained by transforming the integral equation (\ref{eq:MonoD1d}) to a differential equation
\begin{equation}
(\nabla^2+k^2)\langle{\tilde D}^+\rangle =
\nabla^2\langle{ P}^+\rangle\,,
\label{eq:FEX}
\end{equation}
where the polarization density is
\beq
{ P}^+ (x) = \sum_j \mathfrak{ P}^{(j)}\delta(x-x_j)\,.
\eeq
The model can incorporate different forms of the incident field, but in the present work we concentrate on the plane wave
\beq
\tilde D^+_F(x)=D_0 \exp (ikx)\,.
\eeq

The recurrent scattering in the waveguide can induce strong correlations between the atoms. Nevertheless, for two-level atoms in the limit of low light intensity, these can be entirely understood by classical coupled-dipole model of electrodynamics~\cite{Ruostekoski_waveguide,Javanainen1999a}.  The optical response of the atoms in the low-light-intensity limit can be calculated for a discrete set of atomic positions $\{x_1,x_2,\ldots,x_N\}$ by solving for the coupled set of classical electrodynamics equations for point dipoles and light.
The equations of motion for the dipole moments may be cast in the form
\begin{align}
\dot{\mathfrak{ P}}^{(j)}  = & (i\Delta_j-\gamma_t){\mathfrak{ P}}^{(j)} + \frac{2i\gamma_w}{k} \tilde D^+_F(x_j) \nonumber\\
&-\gamma_w  \sum_{l\neq j} e^{i|x_j-x_l|}\,\mathfrak{ P}^{(l)}\,,
\label{TIMEDEPEQ}
\end{align}
where the radiative linewidth,
\beq
\gamma_t=\gamma_l+\gamma_w\,,
\eeq
depends on the radiative losses out of the waveguide $\gamma_l$ and on the decay rate into the waveguide
\beq
\gamma_w = \frac{k{\cal D}^2 }{ 2\pi \xi_\varrho^2\hbar\eo}\,.
\eeq
The detuning of the incident light from the resonance of the atom $j$ is denoted by $\Delta_j=\Omega-(\omega_0+\delta\omega_j)$, where we have written the atomic transition frequency $\omega_0+\delta\omega_j$ in terms of the average transition frequency $\omega_0$ and the possible small atom-dependent shift $\delta\omega_j$. In the case that all the atomic frequencies are equal we set $\delta\omega_j=0$. The reduced dipole matrix element for the atomic transition is denoted by ${\cal D}$.

In the presence of constant driving we are frequently interested in the steady-state solution of \EQREF{TIMEDEPEQ} that satisfies
\beq
\mathfrak{ P}^{(j)} = \alpha_j \tilde D^+_F(x_j) + \eta_j \sum_{l\neq j} e^{i|x_j-x_l|}\,\mathfrak{ P}^{(l)}\,.
\label{eq:classicaled}
\eeq
Here we have defined a single-atom polarizability in a 1D waveguide as
\beq
\alpha_j=-\frac{2\gamma_w}{ k(\Delta_j+i\gamma_t)}
\eeq
and the parameter
\beq
\eta_j\equiv \frac{i\alpha_j k}{2} = \frac{\gamma_w}{i\Delta_j-\gamma_t}\,.
\label{ETADEF}
\eeq

In Eq.~\eqref{eq:classicaled} each dipole is driven by the incident field and the scattered field from all the other $N-1$ dipoles. Once all $\mathfrak{ P}^{(j)}$ are calculated, the scattered fields may be obtained from Eq.~\eqref{eq:MonoD1d}. If, instead of having fixed atomic positions, we sample the atomic positions from a given distribution, we can simulate the optical responses of trapped atomic ensembles. Provided that  we can synthesize a probabilistic ensemble that corresponds to the position correlations between the atoms in the absence of the driving light, stochastic simulations solve the optical response exactly for stationary atoms in the limit of low light intensity~\cite{Javanainen1999a}.
In each stochastic realization of discrete atomic positions $\{x_1,x_2,\ldots,x_N\}$, we solve for the coupled set of classical electrodynamics equations as in the case of fixed atomic positions
Finally, evaluating the ensemble average over many sets of atomic positions generates the exact solution to the optical response with a given atom statistics~\cite{Javanainen1999a}.

\subsection{Collective excitation eigenmodes}

The coupled system for the atoms and light [\EQREF{TIMEDEPEQ}] may be cast into the form
\beq
\dot{{\bf b}} = i \mathcal{H}{\bf b} + {\bf F}\,.
\label{eigensystem}
\eeq
where ${\bf b}$ denotes a vector made of the amplitudes $\mathfrak{ P}^{(j)}$ for the $N$ atoms of the system and ${\bf F}$ represents the driving by the incident field~\cite{JenkinsLongPRB}
\beq
{\bf b}
 =
 \begin{bmatrix}
\mathfrak{ P}^{(1)}
 \\
\mathfrak{ P}^{(2)}
 \\
 \vdots
 \\
\mathfrak{ P}^{(N)}
 \end{bmatrix}, \quad
 {\bf F}
  =  \frac{2i\gamma_w}{k}
 \begin{bmatrix}
\tilde D^+_F(x_1)
 \\
\tilde D^+_F(x_2)
 \\
 \vdots
 \\
\tilde D^+_F(x_N)
 \end{bmatrix} \,.
\eeq
The diagonal elements of the matrix $\mathcal{H}$ then represent the resonance linewidths and line shifts of the individual atoms and the off-diagonal elements the radiative coupling between the different atoms [from the last term of \EQREF{TIMEDEPEQ}]. The properties of the matrix are sensitive to the interatomic separations (via the interaction terms) and to the detunings of the light from the resonances of the individual atoms.
In this paper, we show how the optical response of the atoms inside the waveguide may be understood and how their collective states can be engineered by analyzing and addressing
the eigenvectors of $\mathcal{H}$.

The matrix $\mathcal{H}$ has $N$ eigenvectors ${\bf v}_j$ defining the cooperative radiative excitation eigenmodes of the system. Although the eigenvectors here form a basis, they are generally not orthogonal, since $\mathcal{H}$ is not Hermitian. Consequently, the eigenvalues are generally complex $\delta_j + i \upsilon_j$ where $\delta_j=\omega_0-\omega_j$ is the shift of the collective mode resonance $\omega_j$ from the average single atom resonance frequency  $\omega_0$ that we have chosen as the reference frequency, and $\upsilon_j$ is the collective radiative resonance linewidth. If $\upsilon_j>\gamma_t$, the mode decays faster, and tends to radiate more, than one atom, so such modes are called superradiant. In the case $\upsilon_j<\gamma_t$ the mode is subradiant.

The non-Hermitian nature of the eigensystem complicates the analysis of the dynamics based on the couplings of the collective eigenmodes to the external driving field~\cite{Facchinetti,Hopkins13}.
However, for the two-level atoms we consider here, the eigensystem for arbitrary interatomic separations is symmetric, and we can impose the biorthogonality condition ${\bf v}_j^T {\bf v}_i=\delta_{ji}$ except for possible zero-binorm states for
which ${\bf v}_j^T {\bf v}_j=0$. In order to introduce a measure for the relative population of a particular eigenmode in an excited atomic ensemble, we define the weight
\beq
L_j= \frac{| {\bf v}_j^T {\bf b}|^2}{\sum_i | {\bf v}_i^T  {\bf b} |^2}
\label{eq:measure}
\eeq
for the eigenvector ${\bf v}_j$ in the state $ {\bf b}$. We also consider special cases of regularly spaced atomic arrays with the lattice constants equal to the integer multiples of half of a wavelength. In that case the eigenvectors may be chosen real, and as such satisfy the conventional orthogonality conditions.

\section{Few-atom systems}
\label{fewatoms}

We begin by illustrating the elementary concepts of 1D electrodynamics in the simplest few-atom cases. In particular, we discuss Fano resonances as a mechanism that may create narrow spectral features in the atomic response.

\subsection{One-atom system}

The coupled dynamics for the atoms and light confined inside the waveguide exhibits characteristic behavior of 1D electrodynamics; for instance, for a resonant excitation the first atom can extinguish all of the light -- a phenomenon that appears to have been first predicted theoretically in Ref.~\cite{Javanainen1999a}, and more recently proposed as a mechanism for optical switches~\cite{Fan1} and transistors~\cite{lukintransistor}. In order to see this, we consider Eqs.~\eqref{eq:classicaled} for $j=1,\ldots,N$, and assume that the atomic positions satisfy $x_j>x_1$ for $j>1$, so that $x_1$ refers to the first atom. The solution $\mathfrak{ P}^{(j)}=0$ for all $j\neq1$ can then be valid whenever $\mathfrak{ P}^{(1)} = \alpha \tilde D^+_F(x_1)$ simultaneously satisfies all the equations. For $j>1$ this is true if $\alpha \tilde D^+_F(x_j) + \alpha G(x_j-x_1)\mathfrak{ P}^{(1)}=0$, indicating that the incident light interferes destructively with the light scattered from the first atom. The single-atom power transmission and reflection coefficients are easily found to be (see also Appendix~\ref{TRFAPP})
\beq
T^{(1)}={(\gamma_t-\gamma_w)^2+\Delta^2\over \gamma_t^2+\Delta^2},\quad R^{(1)}={\gamma_w^2\over \gamma_t^2+\Delta^2}\,.
\label{eq:singletrans}
\eeq
We find that the necessary condition for the destructive interference is that the light is on resonance, $\Delta=0$, and the losses from the waveguide are absent, $\gamma_w=\gamma_t$. In this limit all the light is reflected back by the first atom. The total reflection is a generic phenomenon of 1D  scattering, and is, e.g., the origin of the Tonks gas behavior of impenetrable bosonic atoms in strongly confined 1D traps~\cite{Olshanii}.

\subsection{Two-atom system}

\begin{figure*}[tb]
  \centering
  \includegraphics[width=0.5\columnwidth]{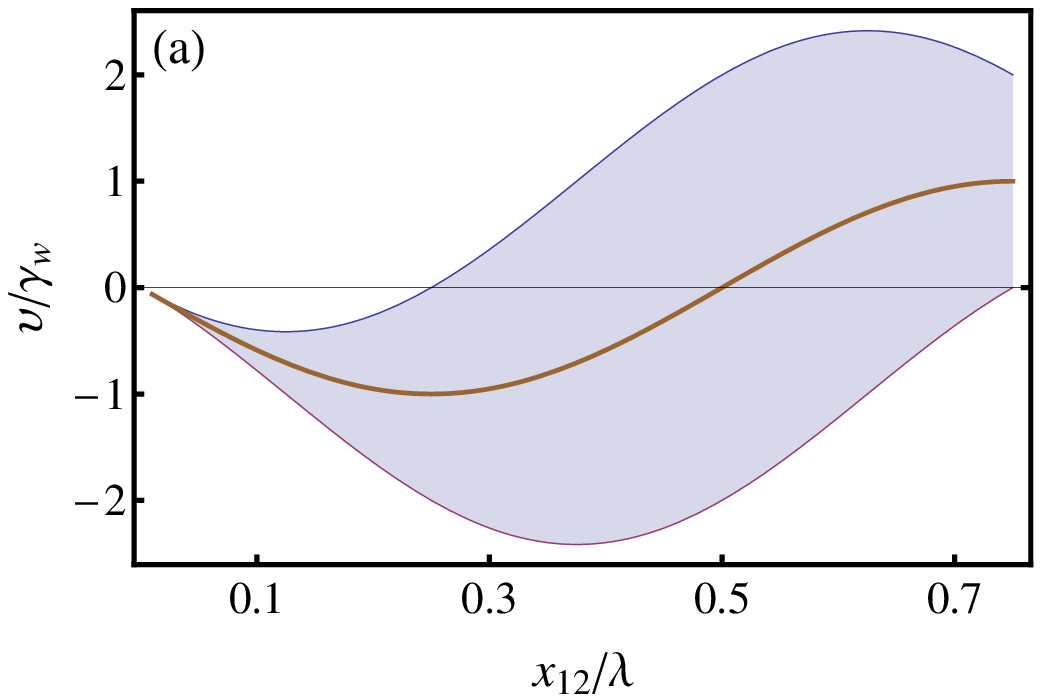}\hspace{-0.3cm}
  \includegraphics[width=0.5\columnwidth]{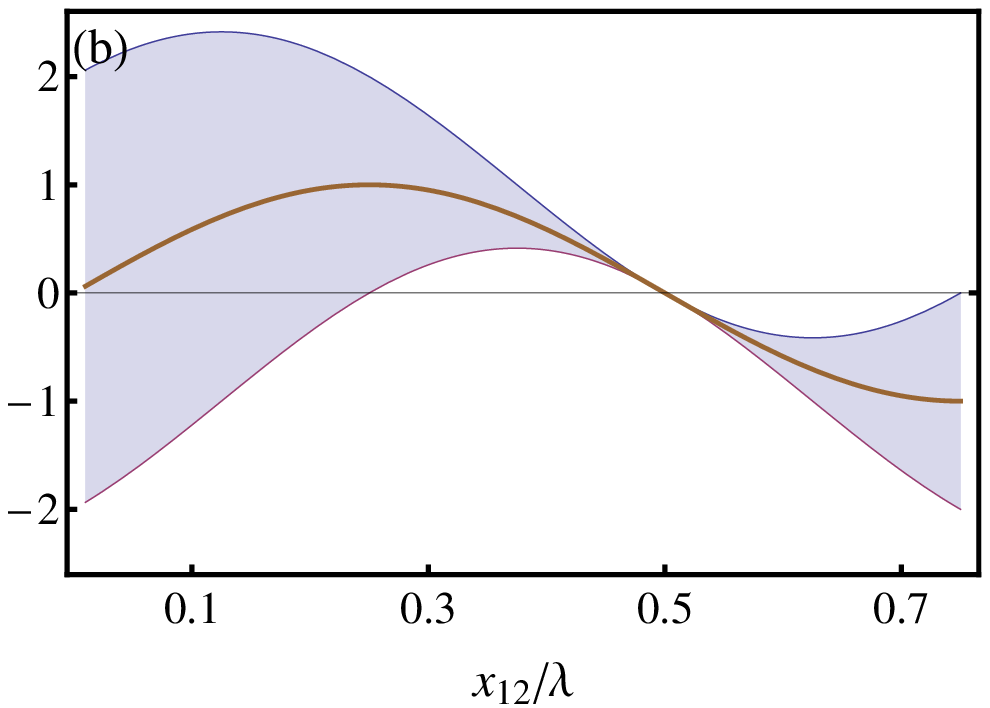}\hspace{-0.1cm}
  \includegraphics[width=0.5\columnwidth]{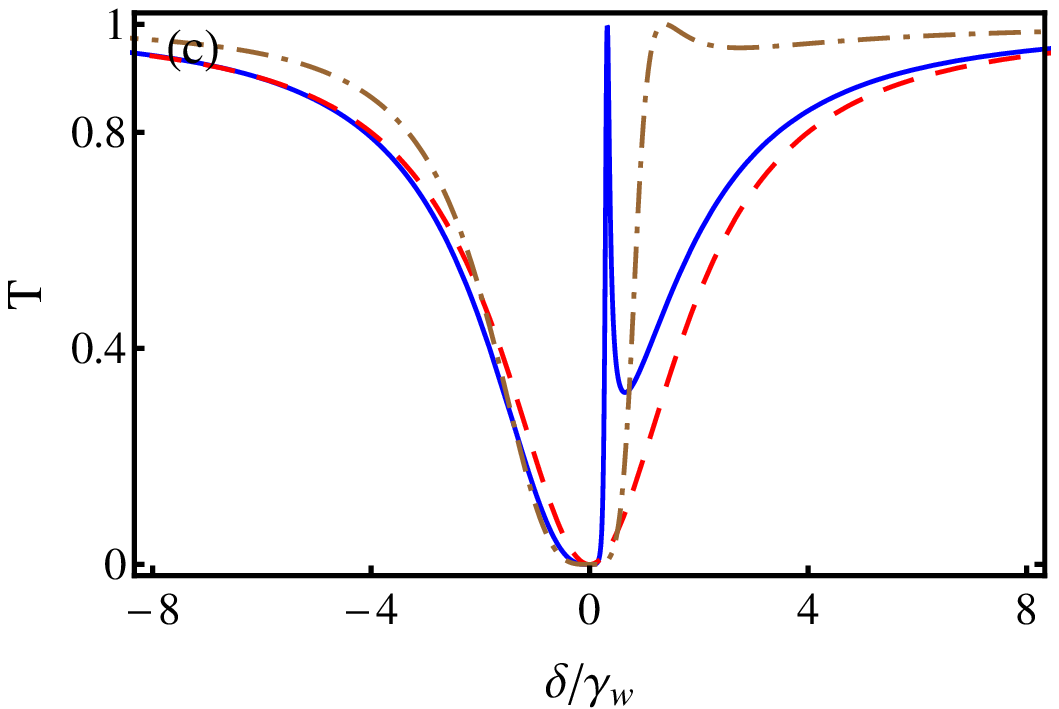}
  \vspace{-0.3cm}
  \caption{Two atoms in a waveguide: The resonance linewidths and line shifts for the collective excitation eigenmodes (a)-(b) and the transmitted total light intensity through the waveguide (c).   The resonance linewidths and line shifts for the antisymmetric (a) and symmetric (b) collective excitation eigenmodes are shown as a function of the interatomic separation. The central line gives the line shift and the shaded region the linewidth. At short interatomic separations $x_{12}<\lambda/4$ the antisymmetric mode is subradiant and the symmetric superradiant. The role of the two modes is interchanged for $\lambda/4<x_{12}<3\lambda/4$.
  The transmitted light intensity as a function of the detuning of the incident light from the single atoms resonance in (c) is shown for $x_{12}=0.5\lambda$ (dashed red), $0.45\lambda$ (solid blue), and $0.35\lambda$ (dash-dotted brown). The broad resonance of the superradiant mode leads to suppressed transmission, except for Fano resonances where the subradiant and superradiant mode interfere destructively. The separation $x_{12}=0.5\lambda$ leads to the excitation of only the antisymmetric, superradiant mode. The case $x_{12}=0.45\lambda$ displays a narrow Fano transmission resonance. The asymmetry for the case $x_{12}=0.35\lambda$ is due to the subradiant mode being broader than at $x_{12}=0.45\lambda$, as shown in (b). This figure is for $\gamma_t=\gamma_w$.
  }
  \label{figtwoatoms}
\end{figure*}

The simplest system to illustrate the cooperative atom response in a waveguide is that of two
atoms. The qualitative features of the light transmission may generally be understood in terms of the collective eigenmodes of the atomic excitations. In the case of two atoms the symmetry of the problem dictates that one mode is symmetric with the same dipole amplitudes, and the other one is antisymmetric with opposite dipole amplitudes.

 In Fig.~\ref{figtwoatoms} we show the resonance linewidths and line shifts for the two eigenmodes, and the transmitted light intensity through the waveguide for a few example cases. The antisymmetric excitation mode is shown in Fig.~\ref{figtwoatoms}(a) and the symmetric in Fig.~\ref{figtwoatoms}(b). At the interatomic separations $x_{12}<\lambda/4$ the antisymmetric mode is subradiant and the symmetric one is superradiant, but the role of the two modes is interchanged for $\lambda/4<x_{12}<3\lambda/4$. As we will see later on in the steady-state response of the more general case of a regular lattice,  the spacing $x_{12}=\lambda/2$ leads to the perfect excitation of only the superradiant mode that is the antisymmetric combination of the two-atom excitations, a classical analogy to the singlet spin state 
\beq
|{\rm sg} \> = \frac{1}{\sqrt{2}} ( |1,g;2,e\> - |1,e;2,g\>)\,.
\label{ANTISSTATE}
\eeq
The periodic arrangement of atoms with a half-a-wavelength separation can therefore be an efficient method of preparing antisymmetric collective excitations for the atoms.

Although a single-photon excitation $|{\rm sg} \>$ represents an entangled state, its atomic excitation dynamics in terms of the Schr\"odinger amplitudes is identical to that of the coupled classical oscillators of Eqs.~\eqref{TIMEDEPEQ} and~\eqref{eq:classicaled}, driven by a coherent field~\cite{SVI10} (with the differences emerging in the photon statistics of emitted light). This is another manifestation of the exact nature of the classical electrodynamics simulations in the limit of low light intensity where the saturation of the atomic transition vanishes~\cite{Javanainen1999a,Lee16}.

In the transmission spectrum the $x_{12}=\lambda/2$ case  is illustrated by a smooth spectral profile where the subradiant mode is entirely decoupled from the incident light (its linewidth exactly vanishes at $x_{12}=\lambda/2$). 
At the interatomic separation $x_{12}=0.45\lambda$ the broad-resonance superradiant and the narrow-resonance subradiant modes interfere destructively, resulting in a narrow Fano resonance for the transmitted light [Fig.~\ref{figtwoatoms}(c)].
The position of the Fano resonance for $x_{12}=0.45\lambda$ at $\Delta\simeq 0.32 \gamma_w$ could also be deduced directly from the subradiant mode resonance curve of Fig.~\ref{figtwoatoms}(b). The interplay between the broad and narrow resonance eigenmodes is even more clearly illustrated in the following three-atom example.

\subsection{Three-atom system}

\begin{figure*}[tb]
  \centering
 \includegraphics[width=0.51\columnwidth]{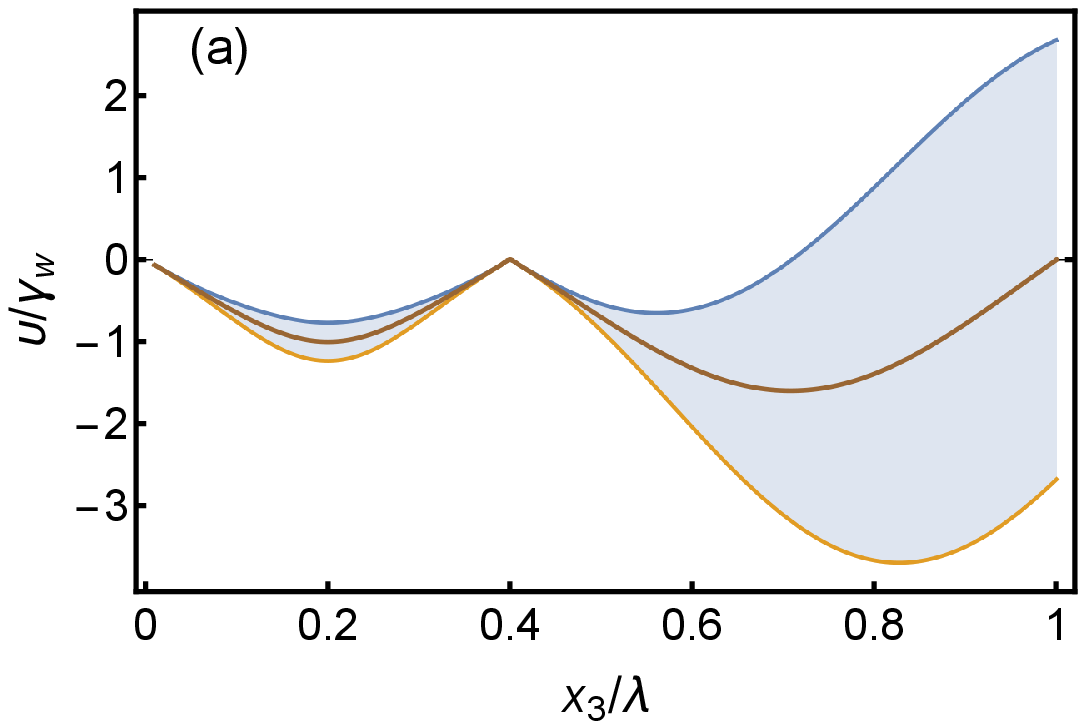}
  \includegraphics[width=0.48\columnwidth]{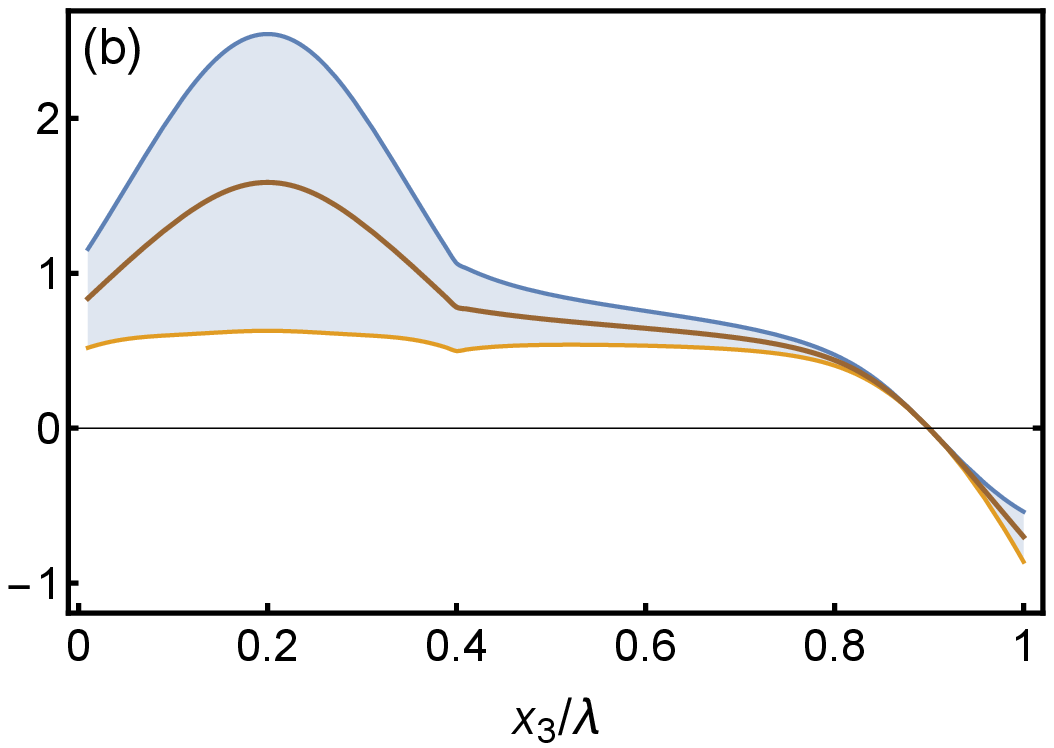}
  \includegraphics[width=0.48\columnwidth]{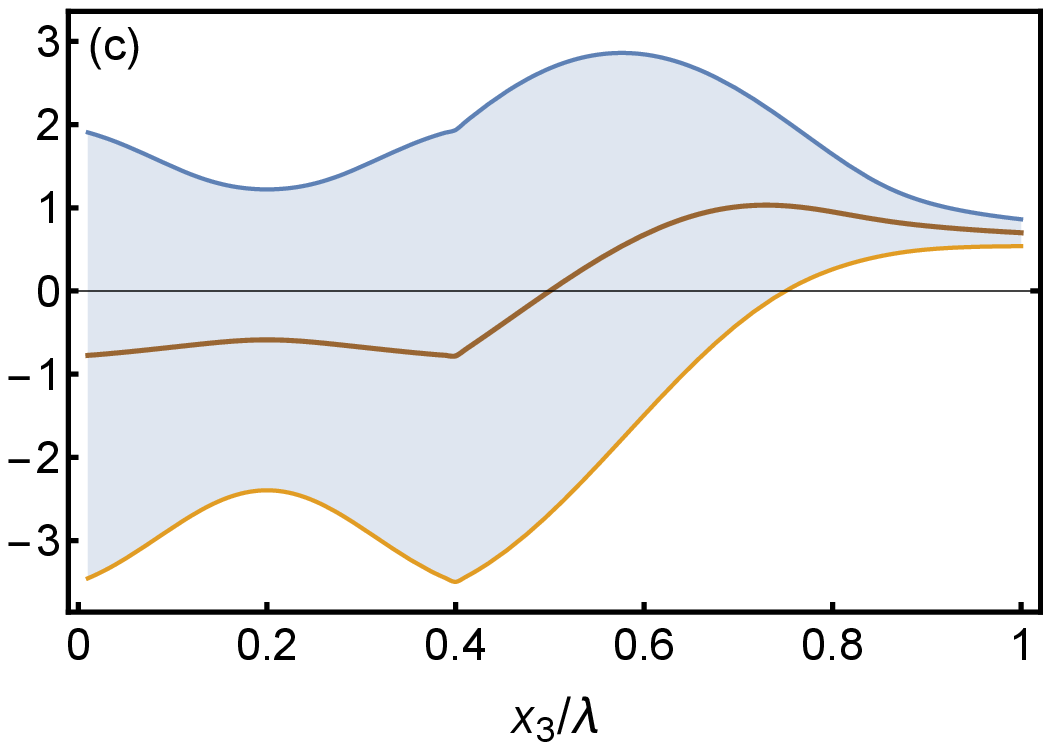}
    \includegraphics[width=0.51\columnwidth]{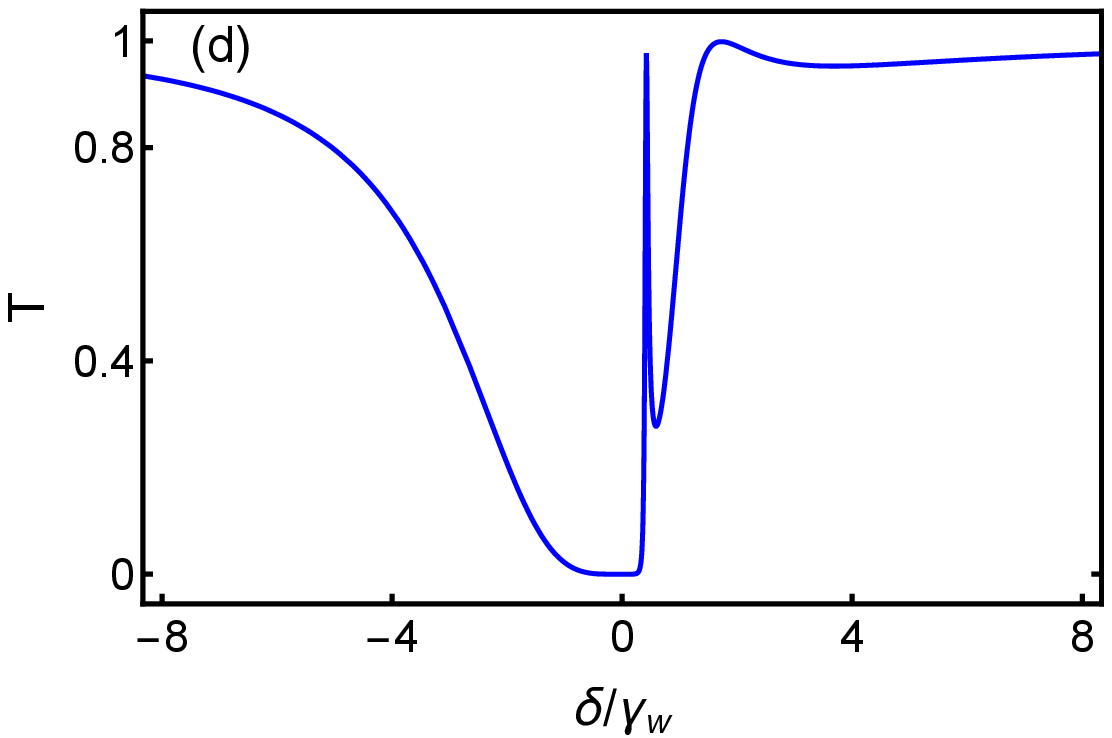}
  \vspace{-0.3cm}
  \caption{The resonance linewidths and line shifts for the collective excitation eigenmodes in a waveguide comprising three atoms and transmitted total light intensity through the waveguide.  (a)-(c)  The resonance linewidths and line shifts for the corresponding collective excitation eigenmodes as a function of $x_3$ when $x_1=0$, $x_2=0.4\lambda$. The central line gives the line shift and the shaded region the linewidth. (d) Transmitted light intensity for three atoms at the locations $x_1=0$, $x_2=0.4\lambda$, and $x_3=0.81\lambda$.
At these values the broad resonance of the superradiant mode in (a) leads to suppressed transmission, except for narrow interferences with the subradiant mode of (b) that results in a Fano transmission resonance. This figure is for $\gamma_t=\gamma_w$.
  }
  \label{figthreeatoms}
\end{figure*}

The role of collective eigenmodes with different resonance linewidths is shown in Fig.~\ref{figthreeatoms}.
We show in Fig.~\ref{figthreeatoms}(a)-(c) the collective radiative resonance linewidths and line shifts for the $N=3$ case when the position of one of the atoms is continuously varied. Depending on the atomic position the subradiant/superradiant character of the modes is strongly affected. The transmission spectrum in the specific example in Fig.~\ref{figthreeatoms}(d) shows strong absorption near the superradiant mode resonance [Fig.~\ref{figthreeatoms}(a) at $x_3=0.81\lambda$]. The transmission also displays a narrow Fano resonance where the superradiant mode interferes with the subradiant mode of Fig.~\ref{figthreeatoms}(b).
It is also easy to demonstrate that the eigenmodes of Fig.~\ref{figthreeatoms}(a) and (b) are complex and not orthogonal at $x_3=0.81\lambda$, even though the biorthogonality condition is satisfied.
The interference with the second subradiant mode [Fig.~\ref{figthreeatoms}(c) at $x_3=0.81\lambda$] leads to a broader transmission peak -- owing to the mode's broader resonance -- that is manifested in the optical response by the strong spectral asymmetry.

 In the Fano resonance, the superradiant mode $|{\rm sp}\>$ can be considered as a `bright' mode that strongly couples to light and the subradiant mode $|{\rm sb}\>$ as a `dark' mode that only weakly radiates. The Fano resonance then has an analogy with the standard single-particle EIT~\cite{FleischhauerEtAlRMP2005}: in the excitation of the superradiant mode $|{\rm sp}\>$ from the initial state $|i\>$ the different paths $|i\> \rightarrow |{\rm sp}\>$ and $|i\> \rightarrow |{\rm sp}\>\rightarrow |{\rm sb}\>\rightarrow |{\rm sp}\>$, etc., via the subradiant mode $|{\rm sb}\>$ destructively interfere. 
Here the effect is inherently a many-body process between the delocalized collective eigenmodes where in this case the coupling between the modes is induced by the nonorthogonality of the eigenmodes.
Analogous light-induced many-body correlated subradiant and superradiant states can be employed for dark-bright mode coupling in plasmonics to produce narrow transmission resonances~\cite{CAIT}.

\section{Periodic lattice of atoms}
\label{arrays}

Our focus is on periodic lattice of atoms. We presently assume that the atoms reside at precisely equidistant positions with the spacing $d$. The configuration approximates, e.g., a very deep optical lattice with the atomic positions fixed at the center of each site $x_j$. Optical tweezers or nanofabrication are two other methods to produce regularly spaced atomic lattices that could potentially be used to achieve very tight spatial confinement of the atoms.

We discuss three aspects of strictly periodic lattices. First, we analyze subradiant and superradiant modes in a lattice with wavelength-related specific spacings between the atoms. Second, we point out narrow spectral features of the lattice, which, we surmise, again are due to Fano resonances between subradiant and superradiant states. Third, we present our finding that under unfavorable conditions the resonance shift of the response of the waveguide may be very large, a circumstance inimical to precision spectroscopy.

\subsection{Atoms separated by a wavelength}

We now investigate the  case when simple analytic considerations are possible, namely, $N$ atoms with the separation of the atoms equal to one wavelength, $d=\lambda$. Interesting and potentially useful connections to steady state may also be made in this case.

\subsubsection{Superradiant and subradiant modes}

The distances between all atoms being integer multiples of the wavelength, the propagation phases $e^{ik(x_i-x_j)}$ are all equal to unity. Moreover, so far we continue with the implicit assumption that the light-atom detuning is the same for all atoms. It is then easy to show that in this special case all the eigenvectors can be chosen to be real. The non-Hermitian eigenvalue problem \EQREF{eigensystem} therefore then unusually exhibits an orthonormal basis of eigenvectors.

It may be seen immediately that a vector ${\bf v}_1$ with all elements equal to one is an eigenvector, and the corresponding eigenvalue $\delta_1+i\upsilon_1$, with $\delta_1=0$ and $\upsilon_1=\gamma_t+(N-1)\gamma_w$, corresponding to a superradiant mode. Writing
\beq
\sum_{l\neq j}\mathfrak{ P}^{(l)} = \sum_{l}\mathfrak{ P}^{(l)} - \mathfrak{ P}^{(j)},
\label{SPLITSUM}
\eeq
one may also see that every vector of polarization amplitudes $\mathfrak{ P}^{(j)}$ with the property $\sum_j\mathfrak{ P}^{(j)}=0$ is an eigenvector with the eigenvalue $i(\gamma_t-\gamma_w)$, representing subradiant modes.
In fact, an arbitrary linear combination of the vectors ${\bf v}_2$, \ldots, ${\bf v}_N$ is a subradiant mode, and the subradiant modes are completely characterized by the condition that  $\sum_j\mathfrak{ P}^{(j)}=0$.

Given that the eigenvector basis in this case may be chosen to be orthonormal, characterization of the steady-state solutions of the optical response [Eq.~\eq{eq:classicaled}] is straightforward.
The incoming plane wave $D_0 e^{ikx}$ has the same value at all atomic sites, the vector made of the inhomogeneous terms in the linear equation is proportional to ${\bf v}_1$, and so is the steady state. The incoming light only couples to the superradiant mode. The steady-state dipole amplitudes are all equal,
\beq
\mathfrak{ P}=\frac{\alpha D_0}{1- (N-1) \eta} e^{ikx_1}\,,
\eeq
and the total transmitted light amplitude reads
\beq
\eo \tilde E^+(x) = D_0 e^{ikx} \big[1+\frac{N\eta}{1- (N-1) \eta} \big]\,,
\eeq
$\eta$ being the value of the parameter~\eq{ETADEF} when all detunings are equal.
As only the superradiant mode is excited, the linewidth of a single atom $\gamma_t$ is broadened to $\gamma_t +(N-1)\gamma_w$ -- and in a lossless waveguide ($\gamma_t =\gamma_w$) to $N \gamma_w$. All the other eigenmodes are subradiant and for a lossless waveguide in fact completely decouple from light, with zero decay rate.

The resonance broadening is due to strong radiative coupling between the atoms. At the resonance we have $\eta=-\gamma_w/\gamma_t$, and for a lossy waveguide $\gamma_t\gg N\gamma_w$, the scattered light amplitude is magnified by the factor of $N$ due to the atom cloud. However, for the case of suppressed losses $\gamma_t\simeq\gamma_w$ no such an amplification occurs and the scattered light can destructively interfere with the incident field leading to strong suppression of multiatom collective scattering, analogously to cold-atom fluorescence experiments in free space~\cite{Pellegrino}.

\subsubsection{Engineering subradiant states and stopping light}

While the subradiant modes do not couple to the incident light and therefore do not get excited, one may nonetheless devise a way to prepare them in steady state and to store the light excitation in an array of atoms. The idea is to relax the condition that the detunings are all equal. In practice, such energy level shifts could be engineered by generating localized AC Stark shifts or magnetic Zeeman shifts.  The overall principle is analogous to the one introduced in Ref.~\cite{Facchinetti} for storing light in subradiant excitations in a 2D array of atoms in free space. Here the waveguide confinement of light makes the effect even more dramatic, and it could potentially be utilized in nanophotonics applications. In the present discussion we again assume a lossless waveguide, $\gamma_t=\gamma_w$.

We thus permit different detunings $\Delta_j$ for the atoms. In this case the steady-state equations (\ref{eq:classicaled}) may be cast in the form
\beq
\mathfrak{ P}^{(j)} = \eta_j K + \eta_j \sum_{l\neq j}\mathfrak{ P}^{(l)}
\eeq
for some constant $K\ne0$. Noting Eq.~\eq{SPLITSUM} and using the definition of $\eta_j$, Eq.~(\ref{ETADEF}), the steady-state equation may be manipulated into the form
\beq
\sum_{l}\mathfrak{ P}^{(l)} = i\left[\sum_i \frac{\gamma_w}{\Delta_i}\right]\left(\sum_{l}\mathfrak{ P}^{(l)}+K\right).
\eeq
For equal detunings a state with $\sum_{l}\mathfrak{ P}^{(l)}=0$ would be subradiant and completely decoupled from light. One may prepare such a state as a steady state for a system with unequal detunings precisely when the inverses of the detunings sum up to zero. Which subradiant states get prepared depends on the values of the individual detunings.

We take as an example a lattice of four atoms. We first calculate the collective excitation eigenmodes when all the detunings of the atoms are equal. The eigenvectors may then be chosen as
\begin{align}
&{\bf v}_1
 = \frac{1}{2}
 \begin{bmatrix}
1
 \\
1
 \\
1
 \\
1
 \end{bmatrix} ,\quad
{\bf v}_2
 = \frac{1}{\sqrt{2}}
 \begin{bmatrix}
-1
 \\
1
 \\
0
 \\
0
 \end{bmatrix}, \nonumber\\
 &{\bf v}_3
 = \frac{1}{\sqrt{2}}
 \begin{bmatrix}
-1
 \\
0
 \\
1
 \\
0
 \end{bmatrix} ,\quad
 {\bf v}_4
 = \frac{1}{\sqrt{2}}
 \begin{bmatrix}
-1
 \\
0
 \\
0
 \\
1
 \end{bmatrix}
 \,\text{.}
 \label{eq:v_vector}
\end{align}
Here ${\bf v}_1$ is a superradiant mode with the linewidth of $4\gamma_w$, and the modes ${\bf v}_2$, ${\bf v}_3$, and ${\bf v}_4$ have vanishing linewidths. The resonance frequency of each mode is equal to that of a single isolated atom. 

We then break the condition that the detunings of all the atoms are equal. This introduces a coupling between the different modes ${\bf v}_j$, as explained in more detail in Appendix~\ref{storageextras}. In order to make a combination of ${\bf v}_2$, ${\bf v}_3$, and ${\bf v}_4$
in the steady state
we pick $\Delta_1=2\gamma_w/3$, and $\Delta_j=-2\gamma_w$ for $j=2,3,4$. This particular choice excites a state where the norms of the the three subradiant modes ${\bf v}_2$, ${\bf v}_3$, and ${\bf v}_4$ are equal, and there is no superradiant component $\propto{\bf v}_1$. The choice of the detunings satisfies the condition that the sum of the inverses of the detunings sum up to zero. After the steady-state solution is reached, we set the detunings of the atoms equal, and the modes ${\bf v}_j$ become the true eigenmodes of the system.

\begin{figure}
  \centering
  \includegraphics[width=0.49\columnwidth]{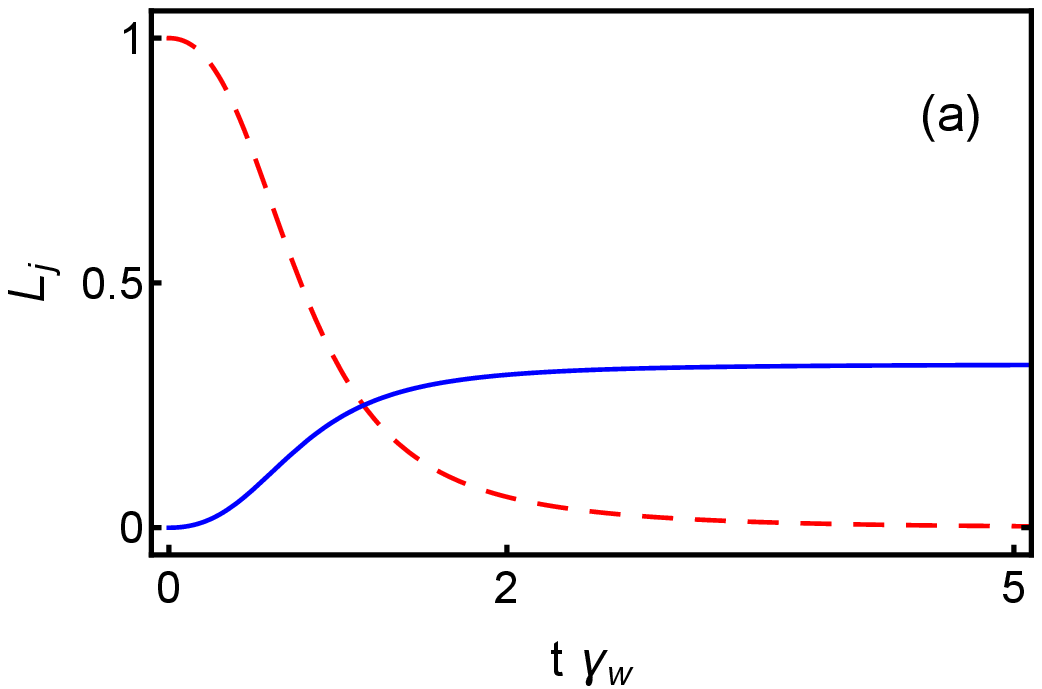}
  \includegraphics[width=0.49\columnwidth]{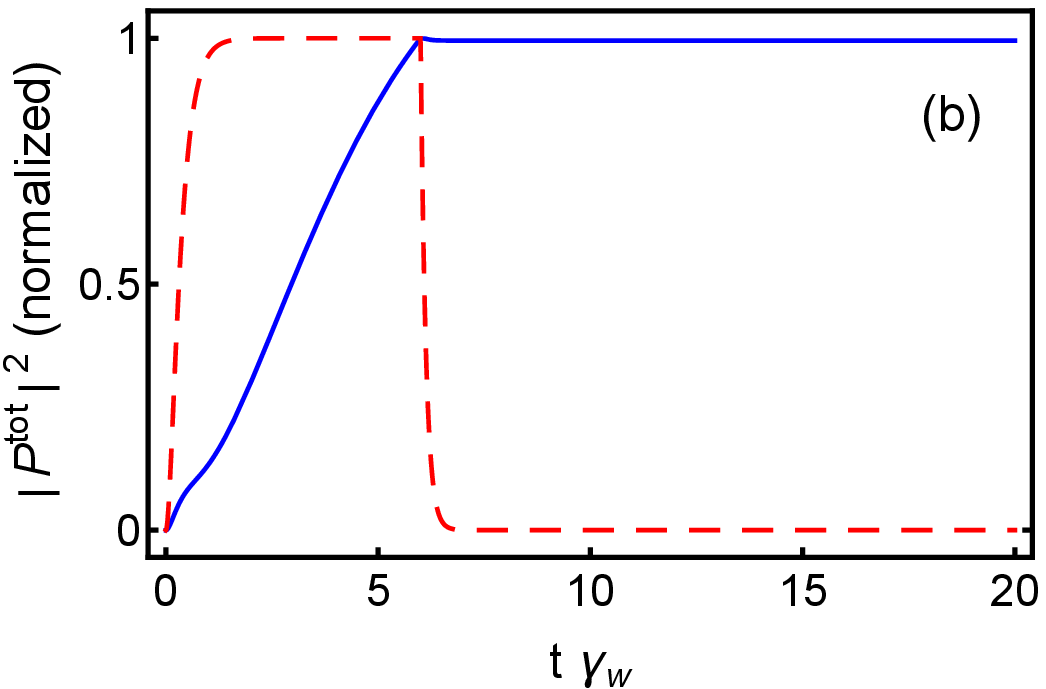}
  \hspace{-0.1cm}
  \vspace{-0.3cm}
  \caption{Engineering collective mode excitations to stop and store light. (a) The populations of the four modes ${\bf v}_j$ ($j=1,\ldots,4$) as a function of time when the system is driven by incident field and the resonance shift of the first atom is different from the rest of the atoms.  The population for the mode $j=1$ (red, dashed line), the modes $j=2,3,4$ are all equal (blue, solid line). The modes ${\bf v}_j$ are eigenmodes only when the shifts of every atom are equal. The differing shifts introduce a coupling between the modes and the excitation is entirely transferred to the modes  ${\bf v}_2$, ${\bf v}_3$, and ${\bf v}_4$. (b) The total atomic polarization density, $\sum_{j} |{\mathfrak{ P}}^{(j)} |^2$,  as a function of time, normalized so that the steady state yields one. We show (blue, solid line) the case where all the excitation is driven to the modes ${\bf v}_2$, ${\bf v}_3$, and ${\bf v}_4$ by the resonance shift, as in (a), that is removed at time $t=6/\gamma_w$ (linearly over the duration of 0.2$\gamma_w$) when the incident field is simultaneously turned off. Now the population is trapped to the modes  ${\bf v}_2$, ${\bf v}_3$, and ${\bf v}_4$ that, in the absence of the resonance shifts, are the eigenmodes with zero resonance linewidths. This is illustrated by the absence of the decay of the excitation. For comparison, we also show the case where the system is driven in the absence of resonance shifts, such that only the superradiant mode ${\bf v}_1$ is excited (red, dashed line), displaying a rapid decay of the excitation. In this figure $\gamma_t=\gamma_w$.
  }
  \label{figsubradiant}
\end{figure}

In Fig.~\ref{figsubradiant}(a) we show the dynamics of the driven four-atom system in the waveguide for our choice of unequal values of the detunings and calculate the populations of each eigenmode using \EQREF{eq:measure}. The population is initially almost solely in the superradiant mode ${\bf v}_1$, but the different resonance shift of the first atom quickly drives the excitations to the modes ${\bf v}_2$, ${\bf v}_3$, and ${\bf v}_4$. In the steady-state the population of the mode ${\bf v}_1$ is, indeed, zero.

In Fig.~\ref{figsubradiant}(b) we show the total atomic polarization density summed over all the atoms, when all the excitation is transferred to the modes ${\bf v}_2$, ${\bf v}_3$, and ${\bf v}_4$ by the shift distribution, similarly as in the case of Fig.~\ref{figsubradiant}(a). At time $t=6/\gamma_w$, the resonance shifts are removed and the incident field is turned off. Now the modes ${\bf v}_j$ of \EQREF{eq:v_vector} are eigenmodes of the system, and all of the excitation is in the modes $j=2,3,4$ that exhibit the zero resonance linewidth. As these modes are entirely decoupled from the radiation field, they cannot decay and the light is therefore stopped and stored in the zero-linewidth subradiant modes.

For comparison, we also show in Fig.~\ref{figsubradiant}(b) the case where only the superradiant mode ${\bf v}_1$ is excited. This happens when all atoms are on resonance. The excitation in this case rapidly decays away after the incident field is turned off.

Our example again illustrates the analogy between the interplay of the subradiant and superradiant collective modes and the dark and bright modes of the single-particle EIT~\cite{FleischhauerEtAlRMP2005}, as also further highlighted in Appendix~\ref{storageextras}. Here the superradiant bright mode of the four-atom system is driven by the incident light. The resonance shift of the first atom plays the role of the coupling field that transfers the population to the dark subradiant mode. By turning off the coupling (here the level shift), the light is trapped to the dark mode that can be utilized in the stoppage and storage of light, analogously to the stopped and stored light of the single-particle EIT~\cite{LiuEtAlNature2001}.

\subsubsection{Atoms with half-wavelength spacing}

Finally we point out that related considerations and results apply also when the spacing of the atoms is equal to half of a wavelength, and even if the array is not periodic as long as the distance between two successive atoms is an integer multiple of half of a wavelength. The latter is the main premise of the present subsection.

Let us assume, without restricting the generality, that the first atom is at $x_1=0$, and define new polarization amplitudes as
\beq
\tilde{\mathfrak{P}}^{(j)} = e^{i k x_j}\mathfrak{P}^{(j)}.
\eeq
The exponential prefactor is always $+1$ or $-1$; $+1$ for the first atom by our convention  $x_1=0$. The effect of the prefactors is that in both the steady state and time dependent equations of motion for the modified amplitudes $\tilde{\mathfrak{P}}^{(j)}$ the exponentials $e^{ik|x_j-x_l|}$, either $+1$ or $-1$, all get replaced by $1$. The mathematics for the modified amplitudes is exactly the same as for the original amplitudes in the case of the separation of one wavelength between the atoms.

As an example, consider the case when the atoms are separated exactly by half a wavelength, whereupon the prefactors alternate between $\pm1$. In terms of the modified amplitudes the superradiant state has all equal polarization amplitudes, which means that the actual amplitudes have equal magnitudes but alternate in sign from one site to the next. This would be an $N$-atom analog of the antisymmetric state~\eq{ANTISSTATE}.

The same scheme also works if the lattice with a half-wavelength spacing has vacancies; the prefactors $\pm1$ simply do not occur in regular alternation. Besides, in the expression for the total field~\eq{eq:MonoD1d} in terms of the modified amplitudes the propagation phases in the Green's function also get modified so that it appears as if the atoms resided at regular one-wavelength spacings. To the outside world there is no difference if the spacing between the atoms is a wavelength, or half of a wavelength, or even irregular as long as the spacing between adjacent atoms is an integer multiple of half of a wavelength.

\subsection{General regular lattice}\label{REGARR}
In the general case of a periodic lattice with fixed atomic positions we obtain analytic solutions for the transmitted light in closed form by employing transfer matrices (see Appendix~\ref{TRFAPP}). One of the features of the transfer-matrix solution is that it depends only on the round-trip phase shifts in light propagation between successive atoms. This tallies with the observations we made in the preceding section about the relations between the results for wavelength and half-wavelength spacings. In the present section we show explicit results only for the case of ideal lossless waveguide, $\gamma_t=\gamma_w$.

\begin{figure}[tb]
  \centering
  \includegraphics[width=0.51\columnwidth]{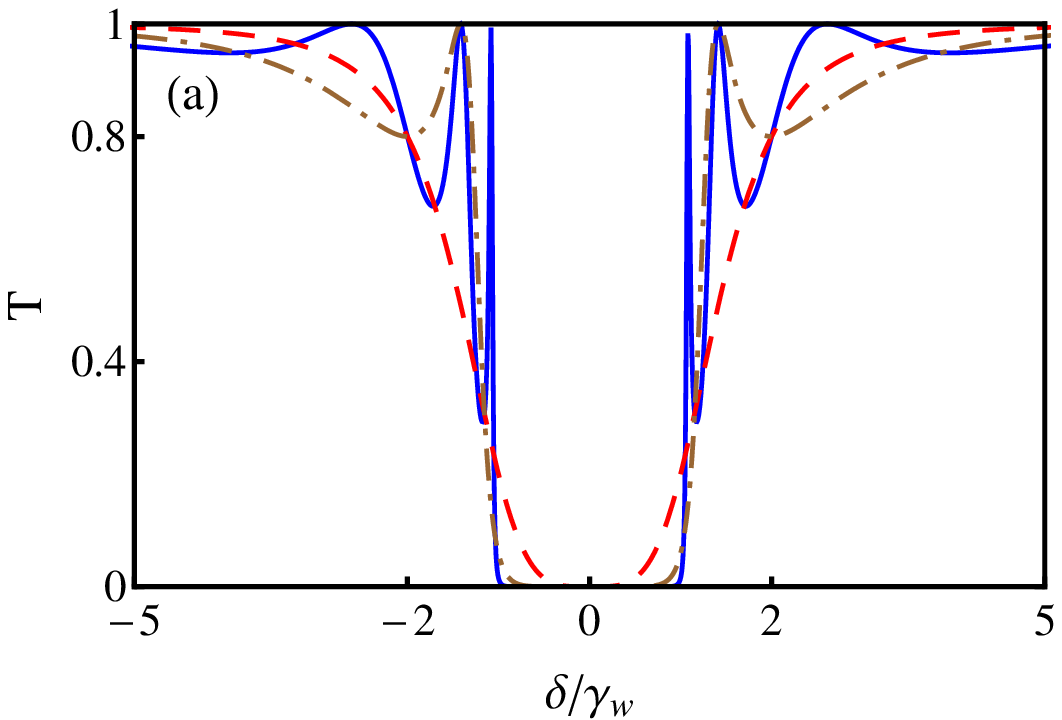}
  \includegraphics[width=0.47\columnwidth]{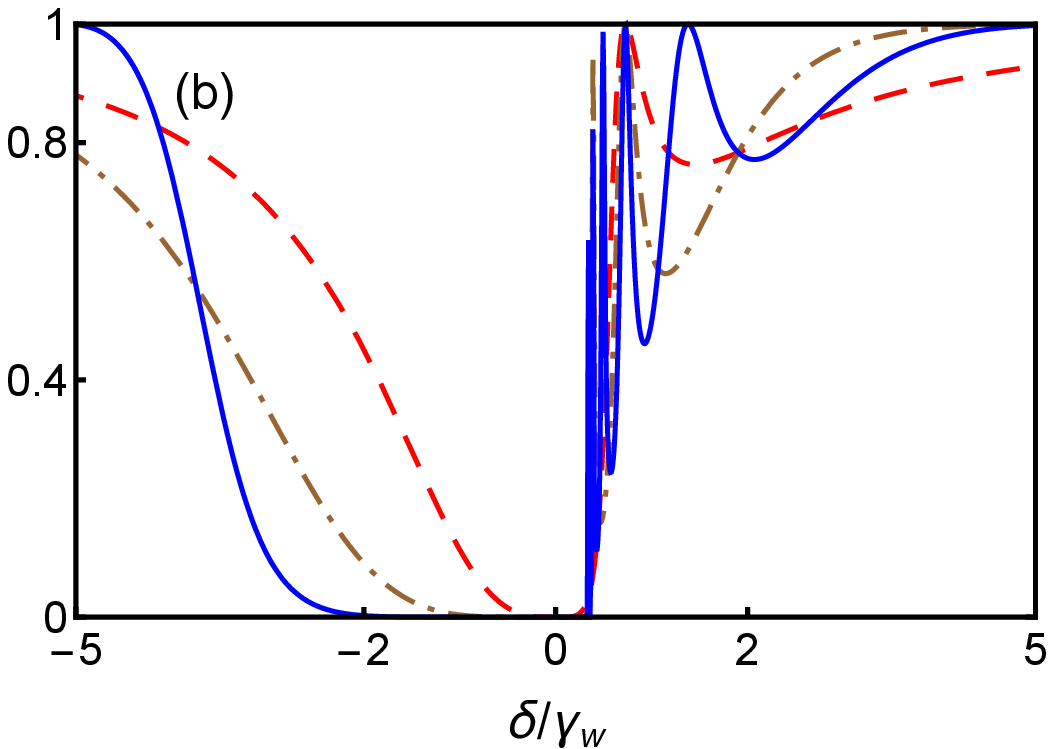}
  \hspace{-0.1cm}
  \vspace{-0.3cm}
  \caption{Transmitted total light intensity through a waveguide with atoms in a periodic lattice of spacing (a) $d=0.25 \lambda$ and (b) $d=0.4 \lambda$.  The transmission coefficient $T$ for 2 (dashed red), 4 (dash-dotted brown), and 8 (solid blue) atoms. The spectra may be analyzed by the closed form analytic solutions. For example, in (a) the $N=8$ case exhibits full transmission with $T=1$ at $\Delta/\gamma_w=\pm [2(2\pm\sqrt{2})]^{1/2},\pm\sqrt{2}$. This figure is for $\gamma_t=\gamma_w$.
  }
  \label{figarray}
\end{figure}

A few example curves for the light transmission through a periodic lattice are shown in Fig.~\ref{figarray} for two lattice spacings. We have discussed the optical response for the case $d=\lambda/2$  already. On the other hand, for the spacing $\lambda/4$ the round trip phase shift between any two atoms is a multiple of $\pi$, and one could anticipate some sort of destructive interference.
This is illustrated in Fig.~\ref{figarray}.
Unlike in the $d=\lambda/2$ case where only the superradiant mode is excited, here subradiant modes also participate, and the lattice response displays a characteristic interference pattern between the modes (as the Fano resonances in the few atom systems in Sec.~\ref{fewatoms}) that becomes rapidly oscillatory as the atom number increases. The analytic solution (see Appendix~\ref{TRFAPP}) provides the values of the detuning for destructive interferences when the waveguide becomes fully transparent.

\subsection{Possibly large line shifts}
We also have  an experimentally relevant and maybe surprising analytical prediction about the behavior of the 1D waveguide: a possibly large line shift when the lattice spacing is {\em close to\/}, but not exactly half of a wavelength.

We start from the exact analytical expressions of the transmission amplitude, (\ref{LATTICERESULT}) in Appendix~\ref{TRFAPP}. In the limit a large fraction of the light leaks out of the waveguide at each atom, small ratio $\gamma_w/\gamma_t$, and for large atom number $N$, we have a simple expansion for the optical thickness
\bea
D^{(N)}&=&-\ln T^{(N)}\nonumber\\
&\simeq&\frac{2 \gamma_t  \gamma_w N}{\gamma_t ^2+\Delta ^2}+
\frac{2 \gamma_w ^2 N \left[\gamma_t ^2+\gamma_t  \Delta  \cot (2kd) -\Delta ^2\right]}{\left(\gamma_t ^2+\Delta
   ^2\right)^2} \,.\nonumber\\
\label{LATTICEAPPR}
\eea
The asymmetry of the second term in the detuning $\Delta$ amounts to a shift of the resonance. The maximum of the resonance line is shifted by
\beq
\Delta_L \simeq  \half\cot(2kd)\,\gamma_w.
\label{LATSHIFT}
\eeq

This resonance shift diverges (and the expansion becomes invalid) when the distance $d$ between the lattice sites tends to (an integer multiple of) half of the wavelength. In fact, we know from our earlier analysis that at $d=\lambda/2$ there is no shift. Nonetheless, it is easy to demonstrate large shifts numerically in the vicinity of $d=\lambda/2$. One way to think about this physically is that each pair of successive atoms defines a cavity. We know that a cavity can pull a resonance; and clearly even more so an array of coupled cavities.

\section{Effects of atomic confinement and photon losses}
\label{confinementlosses}

When the confinement of the atoms is less tight, the fluctuations of the atomic positions affect the transmission of light. As the strength of the confinement varies, the optical response changes.
Here we illustrate how to incorporate such fluctuations for the case of an optical lattice potential.
We consider the lattice in a Mott-insulator state with precisely one atom per site. In an optical lattice the atomic positions in each individual site fluctuate due to the vacuum fluctuations in the ground state. The optical response in such a system could be solved using the Monte-Carlo sampling for atoms in the Mott-insulator state~\cite{Jenkins2012a}.

We consider the atoms in the lowest energy band of a lattice of periodicity $d$  in the axial direction of the waveguide. The vibrational ground-state wave function in each site (the Wannier function) may be approximated by a Gaussian in the axial direction, resulting in the density profile $\rho_j(x)$ with the rms width $\ell=d s^{-1/4}/(\sqrt2\pi)$, where $s$ denotes the lattice depth in the units of the lattice recoil energy $E_r=\pi^2 \hbar^2/(2m d^2)$. Because we have precisely one atom per site, the optical response can then be solved by stochastic simulations in which the position of an atom in each lattice site is sampled independently from the density distribution $\rho_j(x)$~\cite{Jenkins2012a}.

\begin{figure}
\hspace{-5pt}\includegraphics[width=0.49\columnwidth]{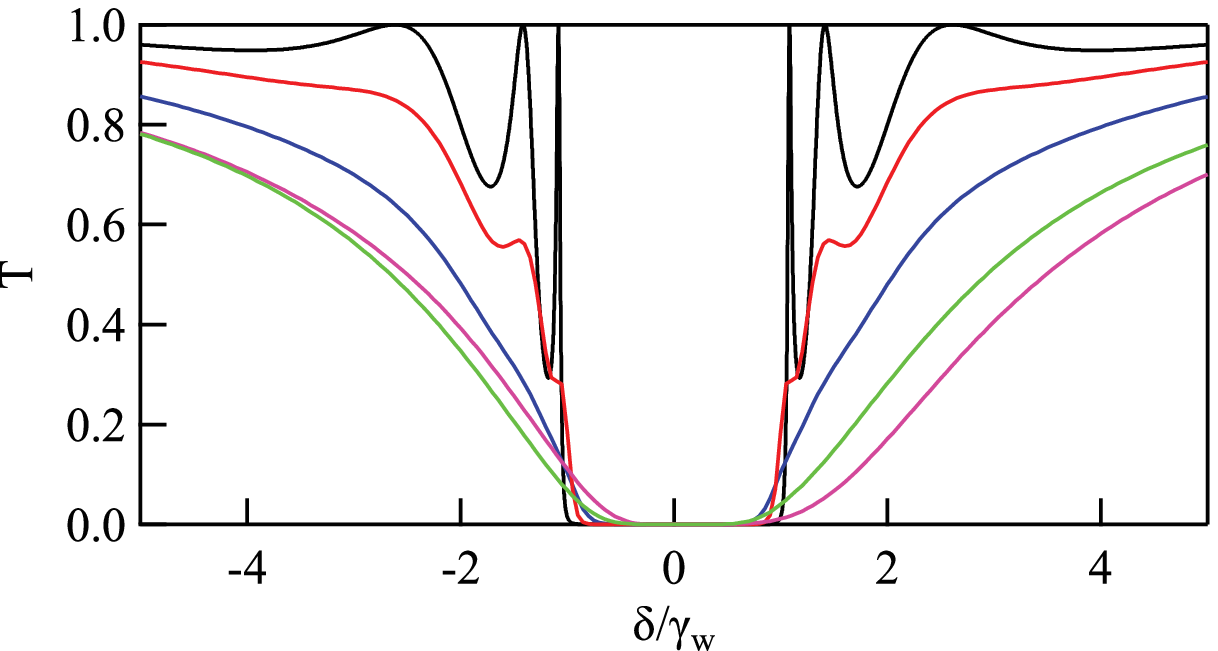}
\includegraphics[width=0.49\columnwidth]{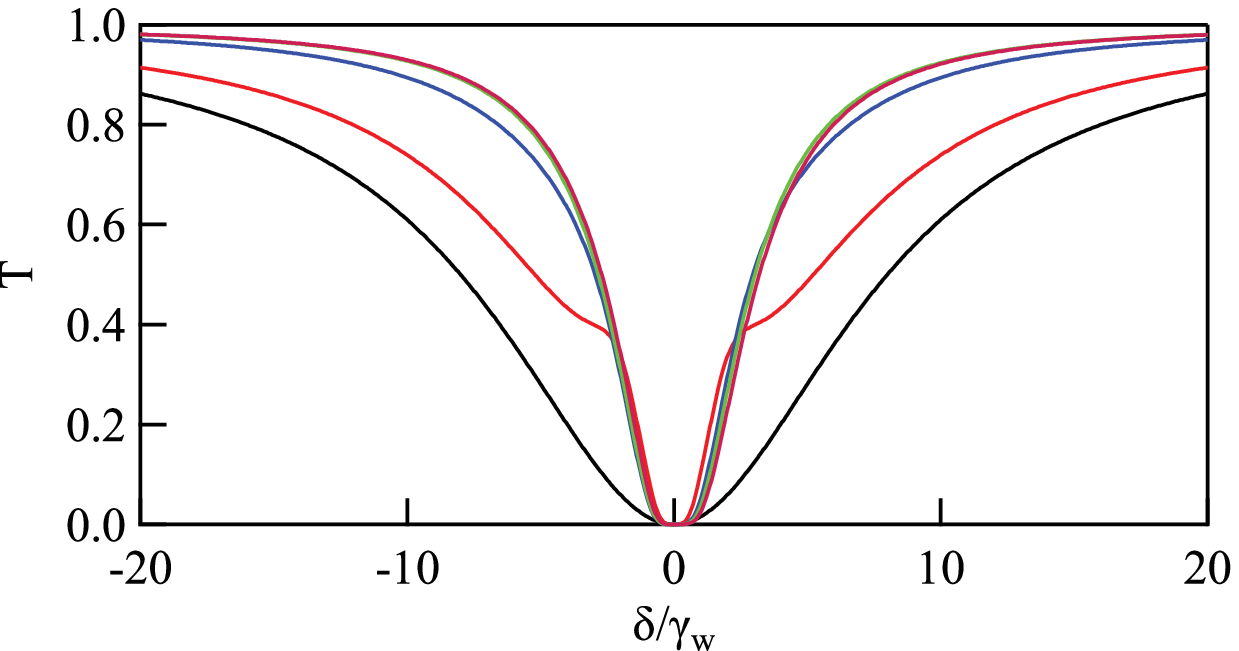}
\caption{The effect of quantum fluctuations in atomic positions on the transmission spectrum of light. The $N=8$ atoms are confined in a periodic optical lattice with precisely one atom per lattice site and the lattice spacing of (a) $d=\lambda/4$; (b) $\lambda/2$.
The curves from top to bottom in (a) refer to the cases of increasing position fluctuations: the fixed atomic positions with no fluctuations (black), Gaussian fluctuations with the rms fluctuations
of $\ell=\lambda/32$ (red), $\lambda/16$ (blue), and $\lambda/8$ (green). The curve at the very bottom (purple) represents the response of the atoms at completely random positions over the interval $2\lambda$.
In (b) the top-to-bottom order of the curves is reversed. This figure is without photon losses, $\gamma_t=\gamma_w$.}
\label{FLUCTLAT4}
\end{figure}

In Fig.~\ref{FLUCTLAT4} we show the transmission spectrum for the cases of increasing position fluctuations in an optical lattice. The oscillatory interference effects of superradiant and subradiant modes are still clearly visible at $\ell=\lambda/32$. In an optical lattice this corresponds to the lattice height of $s\simeq 10 E_r$ for $d=\lambda/4$, and $s\simeq 170 E_r$ for $d=\lambda/2$. As the confinement is weakened, the resonance broadens and the interferences from the subradiant modes are smoothened out in the $d=\lambda/4$ case. For the fixed atomic positions, the $d=\lambda/2$ case is purely superradiant, so the weakened confinement leads to the narrowing of the resonance, as the perfect superradiance is destroyed.

\begin{figure}
\hspace{-10pt}\includegraphics[width=0.53\columnwidth]{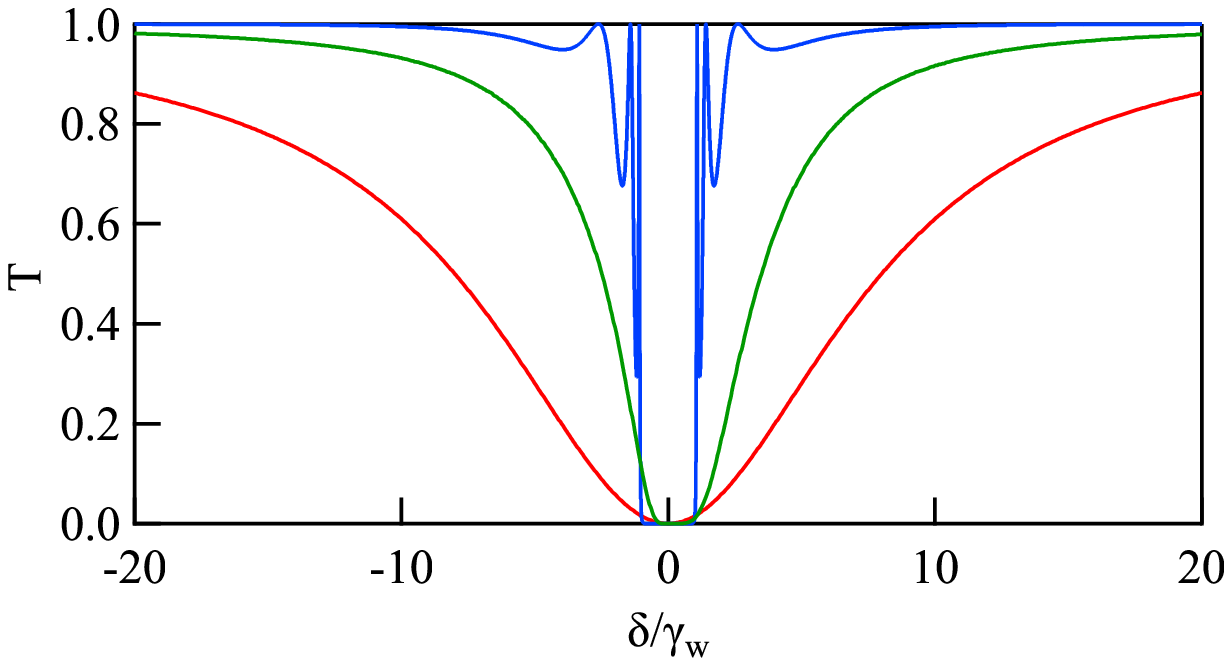}\hspace{-0.5cm}
\includegraphics[width=0.53\columnwidth]{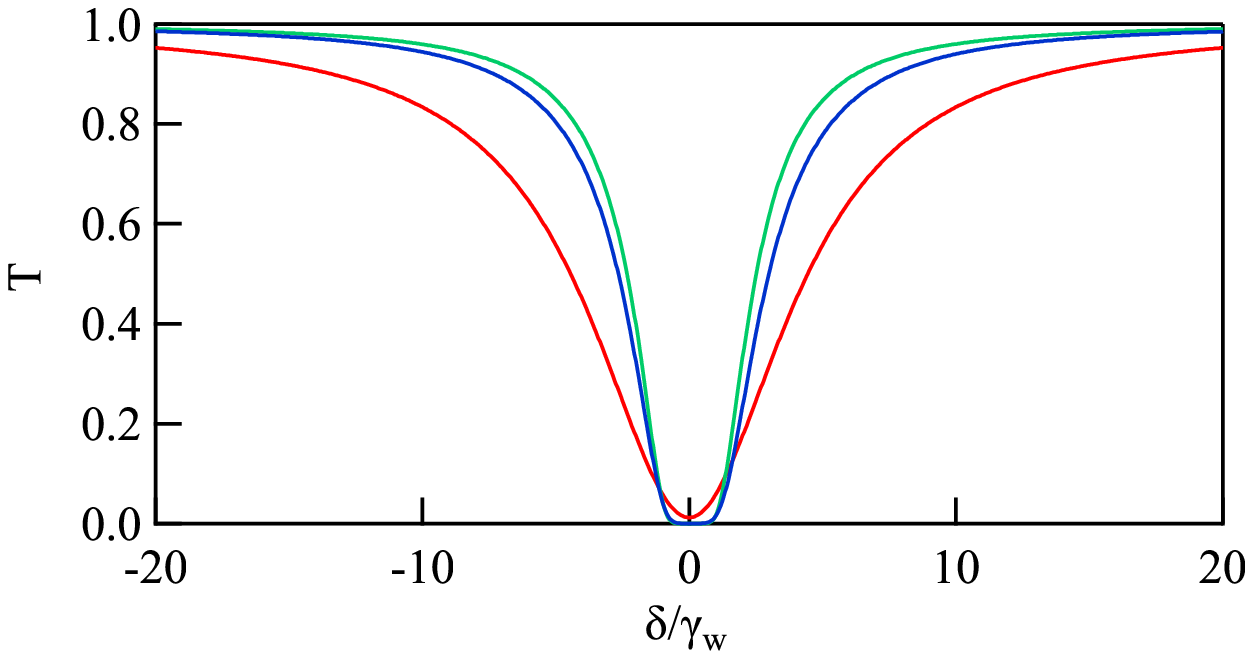}
\caption{The effect of photon losses on the transmission of light through the waveguide. (a) the transmission in the absence of losses for (the curves from the top) $d=\lambda/4$ with fixed atomic positions (blue), entirely random atomic positions (green) over the interval $2\lambda$, and $d=\lambda/2$ with fixed atomic positions (red); (b)  the same with the losses $\gamma_w/\gamma_t=0.5$, resulting in the change in the order of the curves from top to bottom: random positions, $d=\lambda/4$, $d=\lambda/2$. $N=8$ for all graphs. This figure is for strictly fixed atomic positions.}
\label{LATTICE}
\end{figure}

The effect of photon losses from the waveguide is illustrated in Fig.~\ref{LATTICE}. Similarly to the effect of the position uncertainty, the photon losses lead to the broadening of resonance in the $d=\lambda/4$ case with smoothened Fano resonances, and narrowing of the resonance of the superradiant $d=\lambda/2$ case. For comparison we also show the result for random positions of the atoms over the interval $2\lambda$.

\section{Concluding remarks}
\label{conc}

We have demonstrated how the optical response of cold atoms in a waveguide may be understood and described by means of the resonance features of the collective excitation eigenmodes.
Here we have employed the concept of Fano resonance, interference between cooperative modes of the atom-light system, to explain the optical lineshapes of an effectively 1D waveguide. Even if one single atom in a waveguide may be enough to block light propagation completely, in a strictly periodic lattice Fano resonances can also lead to extremely narrow resonances with unit transmission.

The concept of the cooperative eigenmodes is also beneficial in engineering atomic excitations. We showed how the light can be stopped inside the waveguide by driving all the excitation into the zero radiative linewidth collective modes. This can be achieved by dynamically controlling the resonance level shifts of the atoms, so that the light excites a subradiant state that in the presence of equal level shifts for all the atoms would be completely decoupled from the incident light. The proposed technique could be a promising method for preparing many-atom subradiance in waveguides. Although superradiance has been extensively studied in different physical systems~\cite{GrossHarochePhysRep1982}, experiments on many-body subradiance have only been emerging recently~\cite{Guerin_subr16,Jenkins_subr}.

For various reasons, including zero-point motion, the atoms may not have exactly equal spacing; part of the light radiated by the atoms may escape from the 1D waveguide; there may be empty lattice lattice sites; and so on. We have used numerical simulations to study the first two cases. With increasing imperfections, the response of an atomic array to light approaches that of a gas with random positions of the atoms. Rather than this utterly predictable result, our main point here is that the effects of the imperfections may be quantified easily using numerical simulations.

We also have a few analytical observations that might not be so obvious. For instance, the response of a lattice with a strict half-wavelength spacing between the sites is the same for a given number of atoms, no matter which sites the atoms occupy. Moreover, the optical line shape may show a large shift of the resonance if the lattice spacing happens to be close, but not exactly equal to, half of the wavelength of light. This is not as unlikely as it might seem: If resonant light were useful for an optical lattice, the lattice spacing would be exactly half of the wavelength. Evidently, caution should be exercised if one wants to use atomic arrays inside 1D waveguides in precision spectroscopy.

\acknowledgments
We acknowledge support from NSF, Grant No. PHY-1401151, and EPSRC.

\appendix

\section{Transfer matrix solutions}\label{TRFAPP}

\subsection{Basic relations}

For fixed atomic positions the atomic excitations can be solved for the limit of low light-intensity from \EQREF{eq:classicaled}. The total electric field amplitude is then obtained by substituting the solution into \EQREF{eq:MonoD1d}.
In order to analyze the transmission of light by one atom we rewrite Eqs.~\eqref{eq:MonoD1d} and~\eqref{eq:classicaled} in the vicinity of the atom $l=1$ as
\begin{align}
\eo \tilde E^+(x) &= \eo \tilde E_{\rm ext}^+(x)+ {ik\over2}  e^{ik|x-x_1|} \mathfrak{ P}^{(1)},
\label{eq:tr1}\\
\mathfrak{ P}^{(1)} &= \alpha \eo \tilde E_{\rm ext}^+(x_1)\,,\label{eq:tr2}
\end{align}
where $\tilde E_{\rm ext}^+(x_1)$ represents the incident field plus the scattered fields by all the other atoms in the ensemble. We separate the external fields that propagate from the negative ($x_-<x_1$) and positive  ($x_+>x_1$) $x$ directions {\em toward\/} the atom,
\beq
\tilde E_{\rm ext}^+(x_1)= \tilde E_{{\rm ext},-}^+(x_-) e^{ik(x_1-x_-)}+\tilde E_{{\rm ext},+}^+(x_+) e^{-ik(x_1-x_+)}\,.
\eeq
Similarly, we separate the total scattered field $\tilde E^+(x) $ into the field propagating in the positive $x$ direction \emph{away} from the atom, $\tilde E^+_+(x) $, and the field propagating in the negative $x$ direction away from the atom, $\tilde E^+_-(x) $. Substituting these into Eqs.~\eqref{eq:tr1} and~\eqref{eq:tr2} and separating the components of the two propagation directions, we find
\begin{align}
\tilde E^+_+ &= \tilde E_{{\rm ext},-}^+ +\eta (\tilde E_{{\rm ext},+}^+ + \tilde E_{{\rm ext},-}^+),\\
\tilde E^+_- &= \tilde E_{{\rm ext},+}^+ +\eta (\tilde E_{{\rm ext},+}^+ + \tilde E_{{\rm ext},-}^+)\,.
\end{align}
We solve these equations for the field amplitudes for the region $x>x_1$ in terms of the amplitudes for $x<x_1$. The fields at $x_1+$ and $x_1-$ are related by
\begin{align}
&\left[
\begin{array}{c}
\tilde E^+_+ \\
\tilde E_{{\rm ext},+}^+
\end{array}
\right]
=
{\cal T}
\left[
\begin{array}{c}
\tilde E_{{\rm ext},-}^+\\
\tilde E^+_-
\end{array}
\right],\\
&{\cal T} (\Delta) =  \left[
\begin{array}{cc}
 \frac{(2 \eta +1)}{\eta +1} & \frac{ \eta }{\eta +1} \\
 -\frac{\eta }{(\eta +1)} & \frac{1}{(\eta +1)} \label{eq:transfer_singleatom}\\
\end{array}
\right],
\end{align}
where ${\cal T}$ is the transfer matrix for this problem.

For a single atom the transmission and reflection amplitudes,  $t^{(1)}$ and  $r^{(1)}$ respectively, follow directly from the transfer matrix. On the right side of the atom there is only the transmitted wave, call its amplitude $\tilde E^+_t$, whereas on the left we have the incoming and reflected waves, $[\tilde E^+_i,\tilde E^+_r]^T$. These satisfy
\beq
\left[
\begin{array}{c}
\tilde E^+_{t} \\
0
\end{array}
\right]
={\cal T} \left[
\begin{array}{c}
\tilde E^+_{i} \\
\tilde E^+_{r}
\end{array}
\right]\,.
\label{eq:1atomtrans}
\eeq
We obtain
\begin{align}
t^{(1)} (\Delta) &={\tilde E^+_{t}\over \tilde E^+_{i}}= {(\gamma_w-\gamma_t)+i\Delta\over i\Delta-\gamma_t},\\
r^{(1)}(\Delta) & ={\tilde E^+_{r}\over \tilde E^+_{i}} =\eta ={\gamma_w\over i\Delta-\gamma_t}\,,
\label{eq:singletrans_amp}
\end{align}
where we write
\beq
\eta = \sqrt{R^{(1)}} \zeta, \quad \zeta=e^{i\varphi}, \quad \varphi=\arctan(\Delta/\gamma_t)\,.
\label{reflectionstuff}
\eeq
Here $\varphi$ denotes the phase associated with reflection.
The transmission and reflection can be described in terms of the single-atom power transmission and reflection coefficients $T^{(1)} =|t^{(1)} |^2$ and $R^{(1)}= |r^{(1)} |^2$, given in \EQREF{eq:singletrans}.

For the transfer matrix of two atoms at $x_1$ and $x_2$ we also need to consider the propagation phases of the light from $x'$ to $x$, which are governed by the matrix
\beq
\Phi(x,x')  = \left[\begin{array}{cc}
e^{ik(x-x')} & 0\\
0&e^{-ik(x-x')}
\end{array}\right]
\eeq
for both the right- and left-propagating waves.

When considering amplitude transmission, ordinarily one is comparing the properties of the light between some fixed points before and after the sample, call them $x_0$ and $x_3$. We are then back to the same problem as in the case of a single atom, except that the composite transfer matrix has to include the two atoms and the propagation phases:
\beq
{\cal T}_{03}=\Phi(x_3,x_2) {\cal T}(\Delta_2)\Phi(x_2,x_1){\cal T}(\Delta_1)\Phi(x_1,x_0).
\eeq
We obtain the two-atom transmission amplitude
\beq
t_{12}^{(2)} = { t_2^{(1)} t_1^{(1)} \over 1 - \sqrt{R_1^{(1)} R_2^{(1)}} \zeta_1\zeta_2 \xi^2_{12}} e^{ik(x_3-x_0)}\,.
\label{t12}
\eeq
Here $\xi^2_{12} = \exp(2 i k x_{12})$ (with $x_{12}=x_2-x_1$) is a propagation phase associated with a back-and-forth trip of the light between atoms 1 and 2, and $\zeta_1$ and $\zeta_2$ are phase factors upon reflection from each atom. The overall phase factor for propagation from $x_0$ to $x_3$ is trivial, and is henceforth omitted.

We can write Eq.~\eqref{t12} as a geometric series expansion
\begin{align}
t_{12}^{(2)} &=  t_2^{(1)} t_1^{(1)}  + t_2^{(1)} \sqrt{R_1^{(1)} R_2^{(1)}}  e^{i\phi} t_1^{(1)} \nonumber\\& +   t_2^{(1)} \sqrt{R_1^{(1)} R_2^{(1)}}  e^{i\phi}  \sqrt{R_1^{(1)} R_2^{(1)}}  e^{i\phi} t_1^{(1)} + \ldots \,,
\label{t12exp}
\end{align}
with
\beq
\phi=\varphi_1+\varphi_2+2 k x_{12}\,.
\eeq
The interpretation of Eq.~\eqref{t12exp} is straightforward: the sum is over all the repeated photon exchanges between the two atoms, and each term $(R_1^{(1)} R_2^{(1)})^{1/2}$ represents one recurrent scattering event for the photon.

\subsection{Mean-field theory}

In MFT light-induced correlations are ignored. We may construct the MFT solution from the single atom transmission amplitude \EQREF{eq:singletrans}. Let us assume that for $N$ atoms each atom in a row simply passes on the same fraction of light. The transmission amplitude would then be
\beq
t^{(N)}_{\rm mft} = [t^{(1)}]^N\,.
\eeq
In \EQREF{t12} this corresponds to the limit where the denominator is negligible, which can occur in randomly distributed atomic ensembles at sufficiently low densities~\cite{Ruostekoski_waveguide}.
The MFT result can dramatically fail when the repeated photon exchanges between the same emitters become important, indicating a strong cooperative response.
The effect is reminiscent of the proposed failure of the MFT results in 3D free-space electrodynamics~\cite{Javanainen2014a,JavanainenMFT}.

\vspace{0pt}
\subsection{Periodic lattice}
\label{analyticsolution}

The effect of $N$ atoms in a row, with identical spacing $d$ and propagation delay from one atom to the next, is given by the transfer matrix ${\cal T}'^N$, where we take the single-atom transfer matrix of
\EQREF{eq:transfer_singleatom} and incorporate the propagation phases $\xi\equiv e^{i\phi}\equiv\exp(ikd)$,
\beq
{\cal T}' =  \left[
\begin{array}{cc}
 e^{i\phi}\frac{(2 \eta +1)}{\eta +1} & e^{i\phi}\frac{ \eta }{\eta +1} \\
 -e^{-i\phi} \frac{\eta }{(\eta +1)} & e^{-i\phi} \frac{1}{(\eta +1)} \\
\end{array}
\right].
\eeq
As before, after the last atom there is only the right-propagating wave, whereas on the left we have both the incoming and reflected waves,
\beq
\left[
\begin{array}{c}
\tilde E^+_{t} \\
0
\end{array}
\right]
={\cal T}'^N \left[
\begin{array}{c}
\tilde E^+_{i} \\
\tilde E^+_{r}
\end{array}
\right]\,.
\eeq
From this equation one obtains for the amplitude transmission amplitude
\beq
t^{(N)} = \frac{\tilde E^+_{t}}{\tilde E^+_{i}}=\frac{1}{\left[\left({\cal T}'^{-1}\right)^N\right]_{11}},
\eeq
the inverse of the upper left element of the $N^{\rm th}$ power of the inverse of the matrix ${\cal T}'$. Mathematica has no particular trouble with the algebra and gives the result
\begin{widetext}
\beq
t^{(N)}=\frac{2^{N+1} {\cal B} (\eta +1)^N \xi^N}{({\cal A}+{\cal B}-1) ({\cal A}-{\cal B}+1)^N+(-{\cal A}+{\cal B}+1) ({\cal A}+{\cal B}+1)^N},
\label{LATTICERESULT}
\eeq
with
\beq
{\cal A} = (2 \eta +1) \xi^2,\quad {\cal B}=\sqrt{-1+\xi^2} \sqrt{-1+(2 \eta +1)^2\xi^2}\,.
\label{TEMPORARIES}
\eeq
\end{widetext}
The $\xi^N$ is the free-propagation phase factor, including a $\xi$ past the last atom.

\section{Dynamics of stopping light}
\label{storageextras}

In this Appendix~\ref{storageextras} we show how the example on stopping of light in Sec.~\ref{arrays}  can be expressed as  an effective two-mode model for the coupling between the superradiant and subradiant eigenmode manifolds.  
Here we have a regular lattice of four atoms with the spacing $\lambda$.
We can rewrite the coupled system \EQREF{eigensystem} for atoms and light also in the form
\beq
\dot{{\bf b}} = i (\mathcal{H}_0 + \delta\mathcal{H}) {\bf b} + {\bf F}\,,
\label{eigensystem2}
\eeq
where $\mathcal{H}_0$ includes all the terms of the Hamiltonian for the radiatively coupled atoms, except the applied resonance shifts  $\Delta_1=2\gamma_w/3$, and $\Delta_j=-2\gamma_w$ for $j=2,3,4$ which are
included in $ \delta\mathcal{H}$. Any excitation amplitude for the atoms ${\bf b}$ may be expanded in terms of the eigenvectors of the matrix $\mathcal{H}_0$,
\beq
{\bf b}= \sum_n c_n {\bf v}_n = \mathcal{V} {\bf c} \,,
\eeq
where the columns of $ \mathcal{V} $ are formed by the eigenvectors ${\bf v}_n$ and ${\bf c}= [c_1,c_2,\ldots,c_N]^T$. Then we can express the dynamics in terms of the eigenmode amplitudes
\beq
\dot{{\bf c}} = i (\Lambda + \bar{\delta\mathcal{H}}) {\bf c} + \bar{{\bf F}}\,,
\label{eigensystem3}
\eeq
where $\bar{\delta\mathcal{H}}= \mathcal{V}^{-1} \delta \mathcal{H} \mathcal{V}$, $\bar{{\bf F}}=\mathcal{V}^{-1} {\bf F}$, and $\Lambda$ represents a matrix with the eigenvalues of  $\mathcal{H}_0$ in the diagonal.

In the example case one of the eigenmodes ${\bf v}_1$ is superradiant and the other ones ${\bf v}_j$, $j=2,3,4$ have the zero resonance linewidth. The external driving $\bar{{\bf F}}$ then only couples to ${\bf v}_1$, but $\bar{\delta\mathcal{H}}$ introduces a coupling between the mode ${\bf v}_1$ and the manifold spanned by the subradiant mode vectors ${\bf v}_j$, $j=2,3,4$. Our choice of the resonance shifts $\Delta_j$ generally determines which combination of the subradiant modes ${\bf v}_j$, $j=2,3,4$ is excited at the steady state. The present values lead to an excitation that is proportional to the normalized eigenvector
${\bf v}_{\rm sub}=({\bf v}_2+{\bf v}_3+{\bf v}_4)/\sqrt{3} $.

We can then construct from \EQREF{eigensystem3} an effective two-mode model between the excitation amplitudes $c_1$ and $c'= [0, {1\over\sqrt{3}}, {1\over\sqrt{3}}, {1\over\sqrt{3}}]  \cdot  {\bf c} $. We obtain 
\begin{align}
\dot{c}_1 &= i (\Delta+\Delta' + i\upsilon_1) c_1 +i\kappa c' +\bar{F}_1,\label{2mode} \\
\dot{c}' &= i \Delta  c' +i\kappa c_1 \,.\label{2modeb}
\end{align} 
Here the coupling between the superradiant eigenmode and the subradiant eigenmodes is generated by $i [1,0,0,0] \bar{\delta\mathcal{H}} {\bf c} = i\Delta' c_1+ i\kappa c'$ and $i [0, {1\over\sqrt{3}}, {1\over\sqrt{3}}, {1\over\sqrt{3}}]  \bar{\delta\mathcal{H}} {\bf c} =  i\kappa c_1$, with $\Delta'=-4\gamma_w/3$ and $\kappa=-2\gamma_w/\sqrt{3}$.
The linewidth in the subradiant manifold vanishes  $\upsilon_j=0$ (for $j=2,3,4$) and $\upsilon_1=4\gamma_w$. 

Similar two-mode model in case of a 2D lattice of atoms in Ref.~\cite{Facchinetti} provided a good qualitative description of the engineering of subradiant excitations even in the case of a large system. The coupling dynamics between the dark subradiant and the bright superradiant modes in Eqs.~\eqref{2mode} and~\eqref{2modeb} closely resembles the single-particle EIT dynamics.

\end{document}